\documentclass[a4paper,11pt]{article}
\usepackage{jheppub}
\usepackage{tikz}
\usepackage{soul}
\usepackage{graphicx}

\usetikzlibrary{positioning,arrows}
\usepgflibrary{arrows.meta}
\usetikzlibrary{decorations.pathmorphing}
\usetikzlibrary{decorations.markings}
\usetikzlibrary{math}

\tikzset{
Wilson/.style={double distance=1.1pt,postaction={decorate}, decoration={markings,mark=at position .1 with {\arrow{Stealth[scale=1]}},mark=at position .98 with {\arrow{Stealth[scale=1]}}}},
Wilson_1/.style={double distance=1.1pt,postaction={decorate}, decoration={markings,mark=at position .08 with {\arrow{Stealth[scale=1]}},mark=at position .985 with {\arrow{Stealth[scale=1]}}}},
Wilson_3/.style={double distance=1.1pt,postaction={decorate}, decoration={markings,mark=at position .1 with {\arrow{Stealth[scale=1]}}}},
Wilson_Prop/.style={double distance=1.1pt,postaction={decorate}, decoration={markings,mark=at position .2 with {\arrow{Stealth[scale=1]}}}},
Wilson_blank/.style={double distance=1.1pt},
Wilson_arrow/.style={double distance=1.1pt,postaction={decorate}, decoration={markings,mark=at position .7 with {\arrow{Stealth[scale=1]}}}},
Scalar/.style={dashed},
Gluon/.style={decorate, draw=black, decoration={coil,aspect=0.5, post length = 2pt, pre length = 2pt, segment length=3pt,amplitude=3pt}},
Gluon_1/.style={decorate, draw=black, decoration={coil,aspect=0.5, post length = 0pt, pre length = 0pt, segment length=1.75pt,amplitude=1.5pt}},
Gluon_Vertex/.style={decorate, draw=black, decoration={coil,aspect=0.5, post length = 0pt, pre length = 0pt, segment length=3pt,amplitude=3pt}},
Incoming/.style={dashed, postaction={decorate}, decoration={markings,mark=at position 0.6 with {\arrow{Stealth[scale=1.4,reversed]}}}},
Photon/.style={decorate, draw=red, decoration={snake,segment length=3pt, amplitude=1.5pt}},
QuadProp/.style={}
 }

\tikzset{
    halfarrow/.style={postaction={decorate},
        decoration={markings,mark=at position .5 with
       {\arrow{Stealth[scale=1.2]}}}}}
       
\tikzset{axes_style/.style={->}} 
\tikzset{path_style/.style={halfarrow,blue}} 
\tikzset{help_lines_style/.style={dashed}} 
\tikzset{ana_strct_style/.style={dotted}} 
\tikzset{tick_style/.style={}} 
\tikzset{pointer_style/.style={>={Stealth}}} 

\usepackage[T1]{fontenc} 

\usepackage{amssymb,amsmath}
\usepackage[caption=false]{subfig}

\usepackage{breqn}

\newcommand\eps{\epsilon}

\newcommand\cO{  {\cal O}  }
\newcommand{\bea}{\begin{eqnarray}}
\newcommand{\eea}{\end{eqnarray}}

\newcommand{\nn}{\nonumber}

\def\be{\begin{equation}}
\def\ee{\end{equation}}

\newcommand\re[1]{(\ref{#1})}
\def \tr {\mathop{\rm tr}\nolimits}
\newcommand\lr[1]{{\left({#1}\right)}}
\newcommand \vev [1] {\langle{#1}\rangle}

\newcommand\Nc{N}
\newcommand{\Otw}{\mathcal{O}}

\newcommand{\beq}{\begin{equation}}
\newcommand{\eeq}{\end{equation}}

\allowdisplaybreaks[4]

\usepackage{tikz}
\usetikzlibrary{decorations.markings}
\usepgflibrary{arrows.meta}

\tikzset{
    halfarrow/.style={postaction={decorate},
        decoration={markings,mark=at position .5 with
       {\arrow{Stealth[scale=1.2]}}}}}
       
\tikzset{axes_style/.style={->}} 
\tikzset{path_style/.style={halfarrow,blue}} 
\tikzset{help_lines_style/.style={dashed}} 
\tikzset{ana_strct_style/.style={dotted}} 
\tikzset{tick_style/.style={}} 
\tikzset{pointer_style/.style={>={Stealth}}} 
\preprint{IPhT-T19/155, MIT-CTP/5160, MPP-2019-233}
\title{
The full four-loop cusp anomalous dimension in $\mathcal{N}=4$ super Yang-Mills and QCD 
}

\author[a]{Johannes M. Henn,}
\author[b]{Gregory P. Korchemsky}
\author[c]{and Bernhard Mistlberger}
\affiliation[a]{Max-Planck-Institut f\"ur Physik, Werner-Heisenberg-Institut, 80805 Munich, Germany}
\affiliation[b]{Institute de Physique Th\'eorique\footnote{Unit\'e Mixte de Recherche 3681 du CNRS}, Universit\'e Paris Saclay, CNRS, CEA, 91191 Gif-sur-Yvette, France}
\affiliation[c]{Center for Theoretical Physics, Massachusetts Institute of Technology, Cambridge, MA 20139, USA}
\emailAdd {henn@mpp.mpg.de}
\emailAdd {gregory.korchemsky@ipht.fr}
\emailAdd {bernhard.mistlberger@gmail.com}

\abstract{
We present the complete formula for the cusp anomalous dimension at four loops in QCD and in maximally supersymmetric Yang-Mills.
In the latter theory it is given by
\beq
 {\Gamma}^{\rm}_{\rm cusp}\Big|_{\alpha_s^4} = -\left( \frac{\alpha_s N}{\pi}\right)^4  \left[      \frac{73 \pi^6}{20160} + \frac{ \zeta_{3}^2}{8}  + \frac{1}{N^2} 
 \left( \frac{31\pi^6}{5040}  + \frac{9 \zeta_3^2}{4} \right)  \right]  \,.\nonumber
\eeq
Our approach is based on computing the correlation function of a rectangular light-like Wilson loop with a Lagrangian insertion, normalized by the  expectation value of the Wilson loop.  In maximally supersymmetric Yang-Mills theory,
this ratio is a finite function of a cross-ratio and the coupling constant. We compute it to three loops, including the full colour dependence. 
Integrating over the position of the Lagrangian insertion gives the four-loop Wilson loop. 
We extract its leading divergence, which determines the four-loop cusp anomalous dimension. 
Finally, we employ a supersymmetric decomposition to derive the last missing ingredient in the corresponding QCD result. 
}
\keywords{Scattering amplitudes, Gauge theory, Infrared divergences, Casimir scaling}

\begin{document}

\maketitle
\flushbottom
\newpage

\section{Introduction}

 The cusp anomalous dimension is an important quantity in four-dimensional Yang-Mills theories ranging from QCD to maximally supersymmetric $\mathcal N=4$ Yang-Mills theory. It controls the leading ultraviolet divergences of Wilson loops evaluated along a closed contour in Minkowski space-time containing cusps formed by two light-like tangent vectors \cite{Polyakov:1980ca,Korchemsky:1987wg}. Two examples of such contours relevant for our discussion are a wedge formed by two semi-infinite lines  and a null polygon with edges along different light-cone directions. 

The former example plays an important role in the study of infrared asymptotics of on-shell scattering amplitudes and form factors \cite{Korchemsky:1985xj,Collins:1989gx,Collins:1989bt}. It also naturally appears in the analysis of DGLAP splitting functions in the semi-inclusive limit \cite{Korchemsky:1988si,Korchemsky:1992xv},  and in the resummation of large Sudakov corrections due to soft and collinear emissions \cite{Sterman:1986aj,Catani:1989ne}. In all these cases, the contribution of soft particles is in a one-to-one correspondence with ultraviolet divergences of semi-infinite cusped Wilson loops  \cite{Korchemsky:1985xj}, allowing us to
find the asymptotic behaviour of the corresponding physical quantities in terms of the cusp anomalous dimension \cite{Korchemsky:1993uz}.

The study of infrared and collinear singularities in Yang-Mills theory is of considerable interest (see for example refs.~\cite{Almelid:2015jia,Almelid:2017qju,Becher:2019avh,Caron-Huot:2017zfo}) and the cusp anomalous dimension plays a key role in it. 
In particular, the four-loop cusp anomalous dimension is the only missing contribution to understand the double pole in the dimensional regulator of four-loop scattering amplitudes.
Currently, form factors relevant for the precise determination of hadron collider observables such as the Drell-Yan or Higgs boson production cross section~\cite{Mistlberger:2018etf,Anastasiou:2015ema} are known to third loop order~\cite{Gehrmann:2010ue} and the advent of the next order is apparent~\cite{Lee:2016ixa,Henn:2016men,vonManteuffel:2019wbj,vonManteuffel:2019gpr}. 
Knowing the four-loop cusp anomalous dimension, and consequently more of the singularity structure of these quantities serves as stringent check of such a computation. 
In combination with the understanding of how scattering amplitudes factorise as particles approach kinematic limits the understanding of infrared singularities allows to resum parts of the perturbative expansion to all orders. 
The four-loop cusp anomalous dimension is a necessary ingredient for resummation at next-to-next-to-next-to-leading logarithmic accuracy. 
Currently, such computations exist with numerical approximations for the four-loop cusp anomalous dimension for a variety of observables.
This applications range from the extraction of the strong coupling constant from $e^+\, e^- $ event shapes ~\cite{Abbate:2010xh,Becher:2008cf} to precision observables at hadron colliders like the transverse momentum distribution of electro-weak gauge bosons~\cite{Bizon:2017rah,Bizon:2018foh,Bizon:2019zgf,Chen:2018pzu}.

Furthermore, null polygon Wilson loops and the cusp anomalous dimension are of special interest in $\mathcal N=4$ super Yang-Mills theory (sYM). This theory has a number of remarkable properties, e.g. the celebrated AdS/CFT correspondence, and is believed to be integrable, at least in the planar limit  \cite{Beisert:2010jr}. Integrability has been
successfully exploited to predict the cusp anomalous dimension in planar $\mathcal N=4$ sYM theory for any value of the 't Hooft coupling $\lambda=g^2 N$~\cite{Beisert:2006ez}.
Extending the integrability approach to predicting non-planar corrections to the cusp anomalous dimension is currently under active investigation.

Another remarkable property of $\mathcal N=4$ sYM theory is that light-like Wilson loops describe the asymptotic behaviour of off-shell correlation functions in the  light-like limit, when the operators approach the position of vertices of null polygon, and they are dual in the planar limit to the so-called MHV  on-shell scattering amplitudes. We shall use this relationship below. 

At present, the cusp anomalous dimension is known in full in QCD at three loops \cite{Korchemsky:1987wg,Moch:2004pa}. 
At four loops, it has been computed analytically up to one special term that we specify presently. 
The cusp anomalous dimension is known to have the so-called Casimir scaling up to three loops \cite{Korchemsky:1988si}. 
Namely, the dependence of $\Gamma_{\rm cusp}$ on the representation of the $SU(N)$ gauge group only enters through a quadratic Casimir. Starting from four loops this property is violated due to the appearance of a new $SU(N)$ color factor -- the quartic Casimir built out of two completely symmetric invariant $d-$tensors with all indices contracted \cite{Gatheral:1983cz,Frenkel:1984pz}.  

As mentioned above, in $\mathcal N=4$ sYM theory the cusp anomalous dimension is known in the planar limit to any loop order. The non-planar correction first appears at four loops and it arises precisely from the quartic Casimir mentioned above. At large $N$, it is given by the sum of two terms with one of them suppressed by the factor of $1/N^2$. It is this term that generates a nonplanar correction to the cusp anomalous dimension  in $\mathcal N=4$ sYM theory at four loops.~\footnote{Another class of non-planar corrections in $\mathcal N=4$ sYM theory is generated by instanton effects. The leading instanton correction to the cusp anomalous dimension has been computed in \cite{Korchemsky:2017ttd}.}

The new color factor arises from pure gluonic diagrams. Their contribution to  $\Gamma_{\rm cusp}$ is not sensitive to the matter content of the theory and, therefore, is the same in $\mathcal N=4$ sYM theory and in QCD. Thus, computing the four-loop correction to the cusp anomalous dimension proportional to this color factor in the former theory, and adjusting the color factors, we can predict the analogous contribution in QCD. This is the only missing term in the existing four-loop expression for $\Gamma_{\rm cusp}$ in QCD mentioned above. 

In this paper, we compute for the first time analytically the non-planar correction to the four-loop cusp anomalous dimension,
both in ${\mathcal N}=4$ sYM and in QCD. We do so using an approach based on a relationship of Wilson loops with correlation functions of local operators. It allows us to avoid complicated Feynman graph calculations. We validate our results by comparing them against previously known analytic (in the planar case) and numerical (in the non-planar case) values, finding perfect agreement. 
Our result for the four-loop cusp anomalous dimension in ${\mathcal N}=4$ sYM  in the adjoint representation of  $SU(N)$ is 
\begin{align}\label{resultcuspSYMinto}\notag
 {\Gamma}^{\rm}_{\rm cusp} =&{}  \left( \frac{\alpha_s  N }{ \pi }  \right) - \frac{\pi^2}{12}  \left( \frac{\alpha_s N }{ \pi }\right)^2 + \frac{11 \pi^4}{720}  \left( \frac{\alpha_s  N }{ \pi }\right)^3 
  \\
 &
 -   \left[   \frac{73 \pi^6}{20160} + \frac{ \zeta_{3}^2}{8}  + \frac{1}{N^2} 
 \left( \frac{31\pi^6}{5040}  + \frac{9 \zeta_3^2}{4} \right)  \right] \left( \frac{\alpha_s N  }{ \pi }\right)^4 + \cO(\alpha_s^5) \,,
\end{align}
where $\alpha_s=g^2/(4\pi)$ is the fine structure constant. 
The corresponding QCD result can be found in eqs. (\ref{eq:K123L}) and (\ref{eq:K4L}) below.

The outline of the paper is as follows.
In section \ref{sec_correlators}, we begin by reviewing correlation functions in $\mathcal{N}=4$ super Yang-Mills
and their relation to light-like polygonal Wilson loops. We focus in particular on the case of the latter with
a Lagrangian insertion, and review their relationship to the cusp anomalous dimension.
Finally, we discuss the integrand, $F$, for the three-loop correlation function of the null rectangular Wilson loop with
a Lagrangian insertion.
Section \ref{section:loopintegrals} is dedicated to the analytic calculation of the relevant three-loop Feynman integrals.
After reviewing the definitions in \ref{section:defs}, we explain in section \ref{section:improvedbaikov} our method for choosing a basis of integrals 
possessing simple transcendental weight properties. We make use of a convenient choice of loop variables to simplify this analysis. 
In section \ref{section:de} we apply the differential equations method to compute all basis integrals.
We present the integrated results for $F$ to three loops in section \ref{section:Fintegratedresults}.
In section \ref{section:Lintegration}, we derive general formulae for performing the integration 
over the Lagrangian insertion, to extract the cusp anomalous dimension at the next loop order.
Section \ref{section:cusp} contains the main results of this paper -- the expressions for the four-loop cusp anomalous dimension 
in $\mathcal{N}=4$ super Yang-Mills and in QCD, for an arbitrary representation of the Wilson line.
Finally, we conclude and give an outlook in section \ref{section:conclusion}.

\section{Cusp anomalous dimension from a correlation function}
\label{sec_correlators}

The connection between infrared asymptotics of on-shell scattering amplitudes and form factors and ultraviolet divergences of semi-infinite cusped Wilson loops has been previously used to obtain the cusp anomalous dimension.   
In this paper, we follow another approach to computing the cusp anomalous dimension that relies on the relation between off-shell correlation functions of local operators and light-like Wilson loops. 

\subsection{The quartic Casimir terms in different gauge theories} 

\begin{figure}[t]
\includegraphics[scale=0.9]{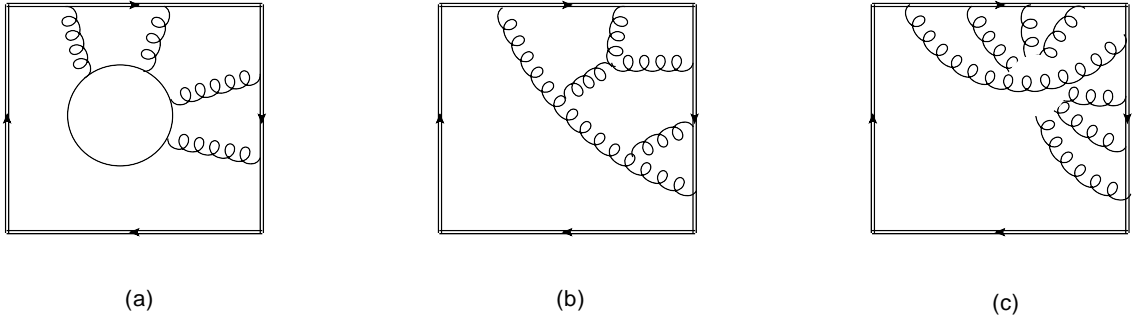}
 \caption{Sample four-loop diagrams contributing to different color structures in eq. (\ref{Gamma-4loops}). Double lines denote the four light-like Wilson lines, and wavy lines represent gluons.  
 Diagram (a) contributes to  $\Gamma_{s}^{(4)}$ or $\Gamma_{f}^{(4)}$ for a scalar or fermion in the loop, respectively; diagram (b) contributes to $\Gamma^{(4)}$; diagram (c) contributes to $\Gamma^{(4)}$ and $\Gamma_{g}^{(4)}$.
 }
 \label{Fig:sample4loopdiagrams}
\end{figure}

As was mentioned above, the cusp anomalous dimension governs the ultraviolet divergences of light-like Wilson loops. 
The simplest example of such an object is a null rectangular Wilson loop 
\begin{align}\label{defcuspedWL}
\vev{W_A(x_1,x_2,x_3,x_4)} = {1\over N_A}\vev{\tr_A P \exp\lr{i\oint_C dx\cdot A(x)} }\,,
\end{align}
where the contour $C$ is a rectangle with vertices located at four points $x_i$ that are light-like separated, $x_{i,i+1}^2 = 0$ (with $i=1,\dots,4$ and $i+4\equiv i$). 
We took the representation to be the adjoint, $R=A$, the reason for such a choice will be clear in a moment. We normalized the Wilson loop by $N_A= \Nc^2-1$, so that its perturbative expansion starts with $1$.

The Wilson loop \re{defcuspedWL} has ultraviolet (UV) divergences due to the presence of the four cusps. In dimensional regularization, with $D=4-2\epsilon$, its leading divergence is \cite{Korchemskaya:1992je,Drummond:2007au}  \footnote{{Here the additional factor of $2$ is inserted to take into account that the Wilson loop is defined in the adjoint representation (and not in the fundamental representation as in the mentioned papers).} \label{foot1}}
\begin{align}\label{W-div}
\log \vev{W_A(x_1,x_2,x_3,x_4)} = - \sum_{L\ge 1} {{2}\over ( L \epsilon)^2}  \lr{\alpha_s \Nc\over \pi}^L \Gamma_{{\rm cusp},A}^{(L)}   + O(1/\epsilon)\,,
\end{align}
where $\Gamma_{{\rm cusp},A}^{(L)}$ are expansion coefficients of the  cusp anomalous dimension in the adjoint representation
\begin{align}
 \Gamma_{{\rm cusp},A} =\sum_{L\ge 1}   \lr{\alpha_s \Nc \over \pi}^L \Gamma_{{\rm cusp},A}^{(L)}\,.
\end{align}
The relation~\eqref{W-div} holds in a conformal Yang-Mills theory, otherwise it is valid up to terms proportional to the beta function.

In a generic Yang-Mills theory with $SU(\Nc)$ gauge group, containing $n_f$ fermions and $n_s$ scalars, the cusp anomalous dimension takes the following general form at four loops
\begin{align} \label{Gamma-4loops}
\Gamma_{{\rm cusp},R}  \big|_{\alpha_s^4} = \lr{\alpha_s \over \pi}^4
 \left[ C_R \Gamma^{(4)} +  {d_R^{abcd} d_A^{abcd}\over N_R}  \Gamma_{g}^{(4)}+  {d_R^{abcd} d_f^{abcd}\over N_R} n_f \Gamma_{f}^{(4)}+{d_R^{abcd} d_s^{abcd}\over N_R} n_s \Gamma_{s}^{(4)} \right],
\end{align}
where 
the subscript $R$ refers to the $SU(\Nc)$ representation in which the Wilson lines are defined, and $N_R=\tr_R1$ is the dimension of the representation.
In the case of QCD and $\mathcal N=4$ sYM the relevant representations are fundamental ($R=F$) and adjoint ($R=A$).
 
The four terms inside the brackets in \re{Gamma-4loops} contain four different color factors depending on $R$. 
Sample diagrams contributing to these terms are shown in Figure~\ref{Fig:sample4loopdiagrams}.
The first term in \re{Gamma-4loops} is proportional to the quadratic Casimir  $C_R = T^a_R T^a_R$.
The corresponding expansion coefficient $\Gamma^{(4)}$ is independent of $R$. 
It only depends on the quadratic Casimir in the adjoint representation, $C_A=\Nc$, as well as on
the number of fermions $(n_f)$ and scalars $(n_s)$  and on the quadratic Casimirs in the representations in which these particles are defined. 
This {proportionality} property is usually referred to as the Casimir scaling.
The coefficient function $\Gamma^{(4)}$ has been computed in refs.~\cite{Henn:2016men,Lee:2016ixa}.

The three remaining terms in \re{Gamma-4loops} involve quartic Casimirs 
built out of completely symmetric tensors 
\beq
d^{abcd}_R =\frac{1}{4!} \sum\limits_{\sigma} \tr(T_R^{\sigma(a)} T_R^{\sigma(b)} T_R^{\sigma(c)} T_R^{\sigma(d)})
\eeq
 in different representations. {The above sum runs over all permutations of  color indices.} Such color factors  appear first at four loops and they violate the Casimir scaling.  The last two terms  in \re{Gamma-4loops} come from Feynman diagrams containing fermion and scalar loops coupled to four gluons. 
The corresponding expansion coefficients $\Gamma_f^{(4)}$ and $\Gamma_s^{(4)}$ have been computed in refs.~\cite{Lee:2019zop,Henn:2019rmi}. 
The main result of this article is the analytic calculation of the last missing coefficient  $ \Gamma_{g}^{(4)}$. 
Numerical results for this quantity were obtained in refs.~\cite{Boels:2017skl,Moch:2017uml,Moch:2018wjh,Henn:2019rmi}.
Since the matter dependence is known, the gluonic coefficient can be computed in maximally supersymmetric Yang-Mills theory.\footnote{This is done using a supersymmetric decomposition, see refs. \cite{Bern:1994zx,Henn:2019rmi}.}
This theory has a number of remarkable properties that simplify the calculation significantly.

\subsection{The cusp anomalous dimension from a finite ratio of Wilson loops} 
\label{sec:cuspfromratioWL}
In principle, one could attempt to compute $\vev{W_A(x_1,x_2,x_3,x_4)}$ in $\mathcal N=4$ sYM directly  in perturbation theory in order to extract $\Gamma_{{\rm cusp},A}$. 
The main complications would be 
a proliferation of Feynman diagrams and the evaluation of the corresponding complicated four-loop Feynman integrals.
 Instead of attempting this direct calculation, we use a method that avoids to a large degree the need for Feynman diagrams, and that only requires the evaluation of a finite three-loop quantity.
 
The key insight is that we do not need to evaluate $\log \vev{W_A(x_1,x_2,x_3,x_4)}$ fully, as the cusp anomalous dimension appears in (\ref{W-div}) as the leading double pole in dimensional regularization. We can imagine $\log \vev{W_A(x_1,x_2,x_3,x_4)}$ as being represented by multiple integrals.
Physically, it is clear that the cusp divergences arise from the integration over particles that propagate at short distances along the light-like edges
adjacent to the cusps. 
The idea is that, upon rescaling the distances, we can isolate a divergent integral over the overall scale and express the cusp anomalous dimension as the finite result of  the remaining integration.
This is similar in spirit to \cite{Erdogan:2011yc}. Here we follow the approach of \cite{Alday:2011ga,Alday:2012hy,Alday:2013ip}.

Let us consider a slightly more general object, namely 
the correlation function of the Wilson loop with the insertion of the Lagrangian normalized by $\vev{W_A}$~\cite{Alday:2011ga,Engelund:2011fg,Engelund:2012re} {(see footnote \ref{foot1})}
\begin{align}\label{F-fun}
{ \vev{W_A(x_1,x_2,x_3,x_4) \mathcal L(x_5)}  \over  \vev{W_A(x_1,x_2,x_3,x_4)} } = {{2}\over \pi^2} {x_{13}^2x_{24}^2\over x_{15}^2x_{25}^2x_{35}^2x_{45}^2} F(x) +O(\epsilon)
\end{align}
It will serve us as the finite integrand mentioned above.
It is possible to show following \cite{Drummond:2007au} that UV divergences cancel in the ratio of correlation functions on the left-hand side of \re{F-fun}, 
so that it remains finite for $\epsilon\to 0$. As a consequence, the first term on the right-hand side  of \re{F-fun} can be defined at $D=4$. 
Conformal symmetry fixes it up to an arbitrary function $F(x)$ of cross-ratio \cite{Alday:2011ga}
\begin{align}\label{x}
x = {x_{25}^2 x_{45}^2 x_{13}^2\over x_{15}^2 x_{35}^2 x_{24}^2}\,.
\end{align}
It depends on four points $x_1,x_2,x_3,x_4$ defining vertices of a null rectangle and the Lagrangian insertion point $x_5$. The invariance of the Wilson loop under cyclic permutation of points $x_1,\dots,x_4$ leads to the relation $F(x)=F(1/x)$. 

In addition, $F(x)$ depends on the rank of the gauge group $N$, and on
the Yang-Mills coupling $g^2$.  As customary in the QCD literature \footnote{Notice that the results in \cite{Alday:2013ip} were expanded in 
powers of $g^2 N/(8\pi^2)$.}, we expand $F(x)$ in powers of the fine structure constant $\alpha_{s}/\pi = g^2/(4\pi^2)$,
\begin{equation}\label{defexpansionF}
\begin{aligned}
F(x) = \left( \frac{\alpha_s N}{\pi}\right)  F^{(0)}(x) 
+ \left( \frac{\alpha_s N}{\pi}\right)^2  F^{(1)}(x) 
+ \left( \frac{\alpha_s N }{\pi}\right)^3  F^{(2)}(x)   \\
+ \left( \frac{\alpha_s N}{\pi}\right)^4 \left[ F^{(3)}_{\rm planar}(x) +\frac{1}{N^2} F^{(3)}_{\rm non-planar}(x) \right] +\cO(\alpha_s^5) \,.
\end{aligned}
\end{equation}
Note that the expansion starts at order $\alpha_s$. The function $F^{(0)}$ defines the correlator \re{F-fun} at Born level and the functions $F^{(L)}$ 
describe corrections at $L$ loops.

Let us now explain how we can combine eqs.~\eqref{F-fun} and~\eqref{W-div} to obtain the cusp anomalous dimension from $F$.
On the one hand, we observe that once we integrate the l.h.s. of eq.~\eqref{F-fun}, this yields a derivative of $\log \vev{W_A(x_1,x_2,x_3,x_4)}$ w.r.t. the coupling. On the other hand, comparing to the r.h.s. of eq.~\eqref{F-fun}, we see that the leading divergence of that quantity is controlled by the cusp anomalous dimension:
\begin{align}\label{integraloverinsertion}
\int \frac{d^{D}x_{5}}{i \pi^{D/2}}  { \vev{W_A(x_1,x_2,x_3,x_4) \mathcal L(x_5)}  \over  \vev{W_A(x_1,x_2,x_3,x_4)} }  =
 - \sum_{L\ge 1}  {{2}\over { L \epsilon^2}}  \lr{\alpha_s \Nc\over \pi}^L \Gamma_{{\rm cusp},A}^{(L)}   + O(1/\epsilon)\,.
\end{align}
As a consequence, once we know the finite ratio in eq.~(\ref{F-fun}) at $(L-1)$ loops, integrating over $x_{5}$ and extracting the leading divergence  allows us to compute the cusp anomalous dimension at $L$ loops.
The detailed derivation was performed in ref.~\cite{Alday:2013ip} using two different regularizations, with the same result.
The effect of the integration over $x_{5}$ is given by a functional $\mathcal{I}$ relating the function $F(x)$ to the cusp anomalous dimension.
We have
\begin{align}\label{extractgammacusp}
\mathcal{I}[F(x)] = -\frac{1}{4} \alpha_{s} \frac{\partial}{\partial \alpha_{s}} {\Gamma}_{\rm cusp}(\alpha_{s}, N) \,,
\end{align}
where $\mathcal{I}$ acts on individual terms as in
\begin{align}\label{rulefunctional-intro}
\mathcal{I}[x^p] = \frac{\sin (\pi p)}{\pi p} \,. 
\end{align}
In order to apply the last relation we write $F(x)$ in the form of a small $x$ expansion. 
The reason why small $x$ asymptotics is related to the cusp anomalous dimension can be understood from eq.~\eqref{integraloverinsertion} - the cusp singularities arise from $x_5$ approaching the cusp points $x_i$. 
Using the definition \re{x}, it is easy to see that in this limit $x\to 0$ or $x\to \infty$. Due to the symmetry $x\to 1/x$ we can map both regions to small $x$.
Eq. (\ref{extractgammacusp}) allows us to determine $\Gamma_{\rm cusp}$ at $L$ loops from $F(x)$ at $(L-1)$ loops, cf. eq. (\ref{defexpansionF}).
This reduces the complexity of the calculation considerably.

\subsection{The Wilson loop integrand from a correlation function} 

Our goal is therefore to obtain the function $F(x)$ at three loops.  
In the conventional Feynman diagram approach one would 
compute  $ \vev{W_A  \mathcal L(x_5)} $ and $\vev{W_A} $  in dimensional regularization  and then find their ratio \re{F-fun}.
Here we follow a different strategy that is based on a relationship between 
correlation functions and light-like Wilson loops. It will allow us to obtain the {\it integrand} for $F(x)$,
bypassing Feynman diagram computations. 

Our starting point is the four-point correlation function of half-BPS operators in $\mathcal N=4$ sYM theory
\begin{align}\label{G4}
G_4=\vev{ \Otw(x_1) \Otw(x_2)\Otw(x_3)\Otw(x_4)}\,,
\end{align}
where $\Otw=\tr(\Phi^I \Phi^J )- \frac16 \delta^{IJ}  \tr(\Phi^K \Phi^K )$ is bilinear in scalar fields $\Phi^I$ (with $I=1,\dots,6$) defined in the adjoint representation of the $SU(\Nc)$.
The operator's scaling dimension $\Delta=2$ is protected from quantum corrections by supersymmetry. 

At weak coupling, the perturbative expansion of $G_4$ in powers of the coupling $\alpha_s$ can be obtained using the method of Lagrangian insertions. 
This method relies on the observation that a derivative of an $n$-point correlation function with respect to the coupling is given by an $(n+1)$-point correlation function involving the insertion of Lagrangian
\begin{align}\label{L-insertion}
\alpha_{s} \frac{\partial}{\partial \alpha_{s}} G_4 = \int d^4 x_5 \vev{\Otw(x_1) \Otw(x_2)\Otw(x_3)\Otw(x_4)\mathcal L(x_5)}\,.
\end{align}
Differentiating repeatedly with respect to the coupling, we can express the $L-$loop correction to $G_4$ in terms of $(4+L)$-point correlation functions involving $L$ insertions of $\mathcal N=4$ sYM Lagrangians
 \begin{align}\notag\label{G4L}
& G_4^{(L)}={1 \over L!} \int d^4 x_5 \dots d^4 x_{4+L} \, G_{4;L}^{(0)}  
\,,
\\
& G_{4,L}^{(0)} = \vev{\Otw(x_1) \Otw(x_2)\Otw(x_3)\Otw(x_4) \mathcal L(x_5) \dots \mathcal L(x_{4+L})}^{(0)} \,.
\end{align}
Here the superscript `$(0)$' was introduced to indicate that the correlation functions are computed in the Born approximation. 

The relation \re{G4L} allows us to interpret $G_{4,L}^{(0)}$ as Feynman integrands at order $g^{2L}$.  
They are rational functions of the distances $x_{ij}^2=(x_i-x_j)^2$ (with $i,j=1,\dots,4+L$).
The form of the functions is heavily constrained by the symmetries of $\mathcal N=4$ sYM theory. 
Namely, the fact that the half-BPS operators $\Otw$ and 
the Lagrangian $\mathcal L$ belong to the same supermultiplet leads to a hidden symmetry of \re{G4L} under permutation of points $x_1,\dots,x_4$ and $x_5,\dots,x_{4+L}$ \cite{Eden:2011we}.
At four loops, this symmetry, combined with conformal symmetry and the requirement for \re{G4L} to have correct behavior in various OPE limits, fixes the leading, planar part of $G_{4,L}^{(0)}$ uniquely whereas the non-planar part of the integrand can be determined up to four arbitrary coefficients \cite{Eden:2012tu}.

These coefficients were determined in the recent paper \cite{Fleury:2019ydf} by matching the non-planar part of $G_{4,L=4}^{(0)}$ 
computed in ref.~\cite{Chicherin:2014uca} using the reformulation of $\mathcal N=4$ sYM in twistor space 
to the analogous expression for the same correlation function obtained in refs.~\cite{Eden:2011we,Eden:2012tu}. Thus, the construction of the four-loop integrand of the correlation function \re{G4} is now completed.

To make a connection to the Wilson loops considered in the previous subsection, we examine the limit $x_{12}^2,x_{23}^2,x_{34}^2,x_{41}^2\to 0$ when four operators in \re{G4} become light-like separated in a sequential manner. As was shown in ref.~\cite{Alday:2010zy}, the correlation function simplifies in this limit and its leading asymptotic behavior is given by the product of Born-level contribution $G_4^{(0)}$ and light-like rectangular Wilson loop $W_A$,
\begin{align}\label{G-W}
\lim_{x_{i,i+1}^2\to 0}G_4/G_4^{(0)} =  \vev{W_A(x_1,x_2,x_3,x_4)}\,.
\end{align}
This relation has a transparent physical meaning. In the first quantized picture, $G_4$ 
describes the propagation of a scalar particle along a closed contour that goes through the point $x_i$. 
In the light-like limit this particle has infinite energy.
 As a consequence, it propagates along a classical trajectory that coincides with $C$ and its interaction with an induced radiation gives rise to an eikonal phase. The latter is given by a Wilson loop in the same representation in which the scalars are defined. This explains the choice of the representation in \re{defcuspedWL}.

We note that eq. (\ref{G-W}) is somewhat formal, as the ratio of four-dimensional correlation functions on the left-hand side
diverges in the light-like limit $x_{i,i+1}^2\to 0$ and requires a regularization.
However, for the purposes of the present paper, we can use the relationship \re{G-W} at the level of the loop {\it integrands}.
In other words, we can profit from the known integrands for the correlation function \re{G4L} in order to obtain the four-dimensional integrand for the function $F(x)$.
In the following, for simplicity of notation, we write the relations between $G_4$ and $F(x)$ in an integral form, but keeping in mind that they hold 
 at the level of the four-dimensional integrands.

Combining together \re{F-fun} and \re{G-W} we obtain
\begin{align}\label{G-F}
{\alpha_{s} \frac{\partial}{\partial \alpha_{s}}}\log (G_4/G_4^{(0)}) = \int {d^4 x_5\over i\pi^2} {{2} \, x_{13}^2x_{24}^2\over x_{15}^2x_{25}^2x_{35}^2x_{45}^2} F(x) + O(x_{i,i+1}^2)
\end{align}
Replacing $G_4$ with its expression \re{G4L} we can now construct the function $F(x)$ at three loops.
Indeed, as was shown in \cite{Alday:2010zy}, the ratio of correlation functions take the following form in the light-like limit
\begin{align}\label{GoverG}
G_4/G_4^{(0)} = 1 +   \sum_{L\ge 1} {1\over L!} \lr{\alpha_s \Nc\over \pi}^L 
\int {d^4 x_5\over i\pi^2}\dots {d^4 x_{4+L} \over i\pi^2}\,
I_L (x_5,\dots,x_{4+L})\,,
\end{align}
where the integrand $I_L$ is obtained from  $G_{4,L}^{(0)}$ defined in \re{G4L} for $x_{12}^2=x_{23}^2=x_{34}^2=x_{41}^2=0$. The explicit 
expressions for $I_L$ are given below.  
Applying \re{GoverG} we obtain an analogous representation for a logarithm of the ratio of correlation functions
\begin{align}\label{logGoverG}
\log(G_4/G_4^{(0)}) =    \sum_{L\ge 1} {1\over L!} \lr{\alpha_s \Nc\over \pi}^L 
\int {d^4 x_5\over i\pi^2}\dots {d^4 x_{4+L} \over i\pi^2} \,
{\mathcal I}_L (x_5,\dots,x_{4+L})\,,
\end{align}
where $\mathcal I_L$ are expressed in terms of the functions $I_L$ according to
\begin{align}\notag
 {\mathcal I}_1 &= I_1(x_5) \,,\qquad
\\\notag
 {\mathcal I}_2  &= I_2(x_5,x_6)- I_1(x_5)I_1(x_6)
\\\notag
 {\mathcal I}_3   &= I_3(x_5,x_6,x_7)+ 2 I_1(x_5)I_1(x_6)I_1(x_7)   
\\
 & - I_1(x_5)I_2(x_6,x_7)- I_1(x_6)I_2(x_7,x_5)- I_1(x_7)I_2(x_5,x_6) \,, 
\end{align}
and so on.
Matching eqs. \re{G-F} and \re{logGoverG} we obtain an integral representation for $F(x)$ in terms of functions $I_L$.
The latter are given by \footnote{{The additional factor of $(-4)^L$ in the denominator comes from the different definition of the integration measure 
in \cite{Eden:2012tu}.}}
\begin{align}
  I_L =   {2 x_{13}^2 x_{24}^2 \over (-4 )^L\prod_{5\le i\le 4+L} x_{1i}^2 x_{2i}^2 x_{3i}^2 x_{4i}^2 \prod _{5\le i<j\le 4+L} x_{ij}^2} P^{(L)}(x_1,\dots,x_{4+L})\,.
\end{align}   
Here $P^{(L)}$ are homogeneous polynomials in $x^2_{ij}$ of degree $(L -1)(L + 4)/2$ whose explicit expressions can be found in \cite{Eden:2011we,Eden:2012tu}. 

 Up to three loops, i.e. for $L\le 3$, the polynomials $P^{(L)}$ do not depend on $\Nc$. At four loops, $P^{(4)}$ receives a non-planar correction
\begin{align}\label{P-nonplanar}
P^{(4)} = P_{\rm planar}^{(4)} + {1\over \Nc^2} P_{\rm non-planar}^{(4)}\,.
\end{align}
As was mentioned above, the planar part can be fixed uniquely whereas the general expression for the non-planar part involves a few arbitrary coefficients
\begin{align}\label{P-Grams}
P_{\rm non-planar}^{(4)} = c_1 \, Q_1 + c_2 \, Q_2+c_3 \, Q_3+ c_4 \, Q_4 +  d_1\, R_1+  d_2\, R_2+  d_3 \, R_3 \,,
\end{align}
where the explicit expressions for $Q_i$ and $R_j$ can be found in \cite{Eden:2012tu} (see Eqs.~(5.23) and (5.20) there). Here the
 polynomials $R_j$ are proportional to Gram determinants depending on the points $x_i$ (with $i=1,\dots,8$).  If all vectors $x_i^\mu$ are four-dimensional, the Gram determinants vanish and $P_{\rm non-planar}^{(4)}$ does not depend on the coefficients $d_i$. 
The coefficients $c_i$ have been determined in ref.~\cite{Fleury:2019ydf}, $c_1=c_2=c_3=0$ and $c_4=-6$.
   
Combining together relations \re{G-F} -- \re{P-nonplanar}, we obtain the following expression for the non-planar correction to the function \re{defexpansionF} 
\begin{align}\label{Fintegrandfinal}
F^{(3)}_{\rm non-planar}(x) ={1\over {3!}}   \int {d^4 x_6 \, d^4 x_7 \, d^4 x_8 \over {4} \, (4\pi^2 i)^{3}} {P_{\rm non-planar}^{(4)} (x_1,\dots,x_8) \over \prod_{6\le i\le 8} x_{1i}^2 x_{2i}^2 x_{3i}^2 x_{4i}^2 \prod _{5\le i<j\le 8} x_{ij}^2}  \,.
\end{align}
The planar correction to $F(x)$ admits a similar representation with $P_{\rm non-planar}^{(4)}$ replaced by a lengthy expression involving a linear combination of $P_{\rm planar}^{(4)}$ and the product of lower loop polynomials $P^{(\ell_1)}P^{(\ell_2)}\dots$ subject to $\ell_1+\ell_2+\dots=4$. To save space, we do not present the corresponding expression.

We can use the relation \re{Fintegrandfinal} together with \re{extractgammacusp} to compute the non-planar correction to the cusp anomalous dimension at four loops. Notice that applying \re{extractgammacusp}, we have to treat $x_5^\mu$ as a $D-$dimensional vector. In this case, the Gram determinants do not vanish, $R_j\neq 0$, and may contribute to \re{P-Grams}. As was shown in \cite{Eden:2011we,Eden:2012tu}, the consistency of \re{GoverG} with the OPE expansion of the correlation function \re{G4} for $x_{i,i+1}^2\to 0$ implies that the polynomial $P_{\rm nonplanar}^{(4)}$ has to vanish when one of the integration points $x_i$ (with $i=5,\dots,8$) approaches the edges of light-like rectangle with vertices at the points $x_1,x_2,x_3,x_4$. In $D=4$ dimension, this leads to 
 the general expression \re{P-Grams}. Repeating the same analysis in $D\neq 4$ we find that the above mentioned condition is satisfied provided that 
\begin{align}
 3 c_4 -3 d_1 + 4d_2 = 0\,.
\end{align}
We use this relation and replace the coefficients $c_i$ with their values to get from \re{P-Grams} 
\begin{align}\label{P-fin}
P_{\rm non-planar}^{(4)} =-6 \,Q_3+ \frac92 R_2 +  d_1\, \lr{R_1+ \frac34 R_2} +  d_3 \, R_3 \,,
\end{align} 
with $d_1$ and $d_3$ arbitrary.
The terms proportional to $d_1$ and $d_3 $ satisfy separately the OPE condition mentioned above. As we discussed, this implies
that the regions of loop integration that potentially produce divergences are suppressed. For this reason, we expect the corresponding
terms to have a smooth and vanishing four-dimensional limit. 
 To verify this, we show in the next section that the $d_{1}$ term on the right-hand side of \re{P-fin} produces a vanishing contribution to \re{Fintegrandfinal} and, therefore, can be safely omitted. For the term proportional to $d_3$, we assume the same based on the above argument, and thus set $d_3 =0$ in our calculation.
 In summary, we obtained the loop integrand for $F(x)$ at three loops from the corresponding correlation function $G_{4}$ at four loops.

Before proceeding to computing the Feynman integrals contributing to $F(x)$, 
we make two further simplifications. First, we make use of the conformal invariance of $F(x)$ to send the point $x_{5}$ to infinity.
This leads to the expression $x= x_{13}^2/x_{24}^2$ for the cross-ratio \re{x}, and likewise removes the $x_{5}-$dependence from eq. (\ref{Fintegrandfinal}).
Second, the resulting expressions for the integrals in (\ref{Fintegrandfinal}) depend on four points $x_1,x_2,x_3,x_4$ that are null separated $x_{i,i+1}^2=0$.
Introducing the dual variables $p_i=x_{i+1}-x_i$,
we notice that these integrals resemble momentum integrals contributing to on-shell four-particle scattering amplitudes.
This will be discussed in the next section.

\section{Three-loop  integrals from differential equations}
\label{section:loopintegrals}

In this section we compute the three-loop Feynman integrals contributing to $F(x)$. 

\subsection{Definition of the integral family}
\label{section:defs}

\begin{figure}[t]
\centering
\subfloat[]{
\raisebox{0.12\height}{
\begin{tikzpicture}
\node at (0,-3/2-1/4) {$x_{1}$};
\node at (0,3/2+1/4) {$x_{3}$};
\node at (-2-1/4,0) {$x_{2}$};
\node at (2+1/4,0) {$x_{4}$};
\node at (0,0) {$x_{7}$};
\node at (-1,0) {$x_{6}$};
\node at (1,0) {$x_{8}$};
\node at (-5/2,-5/2+1/5) {$p_{1}$};
\node at (-5/2,5/2-1/5) {$p_{2}$};
\node at (5/2,5/2-1/5) {$p_{3}$};
\node at (5/2,-5/2+1/5) {$p_{4}$};
\draw (-2-1/5,2)--(-3/2,1);
\draw (-3/2,1)--(-1/2,1) node[draw=none,fill=white,font=\tiny,midway] {1};
\draw (-1/2,1)--(1/2,1) node[draw=none,fill=white,font=\tiny,midway] {3};
\draw (1/2,1)--(3/2,1) node[draw=none,fill=white,font=\tiny,midway] {5};
\draw (3/2,1)--(2+1/5,2);
\draw (-3/2,1)--(-3/2,-1)  node[draw=none,fill=white,font=\tiny,midway] {7};
\draw (-1/2,1)--(-1/2,-1)  node[draw=none,fill=white,font=\tiny,midway] {8};
\draw (1/2,1)--(1/2,-1) node[draw=none,fill=white,font=\tiny,midway] {9};
\draw (3/2,1)--(3/2,-1)  node[draw=none,fill=white,font=\tiny,midway] {10};
\draw (-2-1/5,-2)--(-3/2,-1);
\draw (-3/2,-1)--(-1/2,-1)  node[draw=none,fill=white,font=\tiny,midway] {2};
\draw (-1/2,-1)--(1/2,-1) node[draw=none,fill=white,font=\tiny,midway] {4};
\draw (1/2,-1)--(3/2,-1) node[draw=none,fill=white,font=\tiny,midway] {6};
\draw (3/2,-1)--(2+1/5,-2);
\end{tikzpicture}
}
}
\qquad
\qquad
\qquad
\subfloat[]{
\begin{tikzpicture}
\node at (0.518,1.93) {$x_{1}$};
\node at (1.41421, 1.41421) {$x_{2}$};
\node at (1.93185, 0.517638) {$x_{3}$};
\node at (1.93185, -0.517638) {$x_{4}$};
\node at (0.517638, -1.93185) {$x_{2}$};
\node at (1.41421, -1.41421) {$x_{1}$};
\node at (-0.517638, -1.93185) {$x_{3}$};
\node at (-1.41421, -1.41421) {$x_{4}$};
\node at (-1.93185, -0.517638) {$x_{1}$};
\node at (-1.93185, 0.517638) {$x_{2}$};
\node at (-1.41421, 1.41421) {$x_{3}$};
\node at (-0.517638, 1.93185) {$x_{4}$};
\node at (1.45,2.51) {$p_{1}$};
\node at (2.51147, 1.45) {$p_{2}$};
\node at (2.9, 0.) {$p_{3}$};
\node at (2.51147, -1.45) {$p_{4}$};
\node at (1.45, -2.51147) {$p_{1}$};
\node at (0., -2.9) {$p_{2}$};
\node at (-1.45, -2.51147) {$p_{3}$};
\node at (-2.51147, -1.45) {$p_{4}$};
\node at (-2.9, 0.) {$p_{1}$};
\node at (-2.51147, 1.45) {$p_{2}$};
\node at (-1.45, 2.51147) {$p_{3}$};
\node at (0., 2.9) {$p_{4}$};
\draw (0.75,1.29904)--(1.25,2.16506);
\draw (1.29904,0.75)--(2.16506,1.25);
\draw (1.5,0.)--(2.5,0.);
\draw (1.29904,-0.75)--(2.16506,-1.25);
\draw (0.75,-1.29904)--(1.25,-2.16506);
\draw (0.,-1.5)--(0.,-2.5);
\draw (-0.75,-1.29904)--(-1.25,-2.16506);
\draw (-1.29904,-0.75)--(-2.16506,-1.25);
\draw (-1.5,0.)--(-2.5,0.);
\draw (-1.29904,0.75)--(-2.16506,1.25);
\draw (-0.75,1.29904)--(-1.25,2.16506);
\draw (0.,1.5)--(0.,2.5);
\draw (0,0)--(0,1.5);
\draw (0,0)--(1.29904, -0.75);
\draw (0,0)--(-1.29904, -0.75);
\node at (-0.742307, 0.428571) {$x_{6}$}; 
\node at (0.742307, 0.428571) {$x_{7}$}; 
\node at (0, -0.857143) {$x_{8}$}; 
\draw (0,0) circle (1.5);
\node[draw=none,fill=white,font=\tiny] at (0,3/4) {8} ;
\node[draw=none,fill=white,font=\tiny] at (0.649519, -0.375) {9} ;
\node[draw=none,fill=white,font=\tiny] at (-0.649519, -0.375) {15} ;
\node[draw=none,fill=white,font=\tiny] at (1.06066, 1.06066) {12} ;
\node[draw=none,fill=white,font=\tiny] at (1.44889,  0.388229) {3} ;
\node[draw=none,fill=white,font=\tiny] at (1.44889,  -0.388229) {13} ;
\node[draw=none,fill=white,font=\tiny] at (1.06066, -1.06066) {6} ;
\node[draw=none,fill=white,font=\tiny] at (-1.06066, -1.06066) {10} ;
\node[draw=none,fill=white,font=\tiny] at (0.388229, -1.44889) {14} ;
\node[draw=none,fill=white,font=\tiny] at (-0.388229, -1.44889) {5} ;
\node[draw=none,fill=white,font=\tiny] at (-1.06066, 1.06066) {q8} ;
\node[draw=none,fill=white,font=\tiny] at (-1.44889, -0.388229) {2} ;
\node[draw=none,fill=white,font=\tiny] at (-1.44889, 0.388229) {7} ;
\node[draw=none,fill=white,font=\tiny] at (-1.06066, 1.06066) {1} ;
\node[draw=none,fill=white,font=\tiny] at (-0.388229,  1.44889) {11} ;
\node[draw=none,fill=white,font=\tiny] at (0.388229, 1.44889) {4} ;
\end{tikzpicture}
}
\caption{Diagram (a) represents a  
triple ladder diagram contained in integral family (\ref{definitionFamily}) in the dual variables. 
Each line with index $i$, separating two regions labelled by the coordinates $x_{i_1}$ and $x_{i_2}$, denotes a denominator factor $(x_{i_1,i_2}^2)^{a_i}$
Diagram (b) represents the complete integral family (\ref{definitionFamily}) in the dual variables. The  internal lines with indices $i=1,\dots,15$ correspond to the  $15$ propagator factors in   (\ref{definitionFamily}). }
\label{figuregintegrals}
\end{figure}
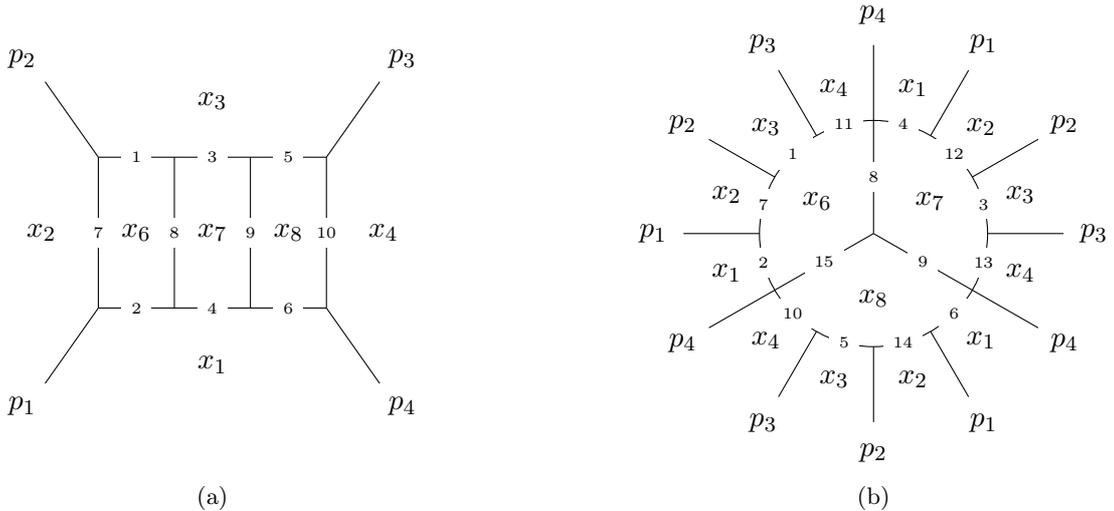

The Feynman integral in \re{Fintegrandfinal} involves the conformal polynomial $P_{\rm non-planar}^{(4)}$ defined in \re{P-fin}. It can be expanded into a sum of terms given the product of distances $x_{ij}^2$. In spite of the fact that the integral \re{Fintegrandfinal} is finite, each individual term may produce divergences. For this reason we  set up the calculation in $D=4-2\eps$ dimensions, and take the limit $\eps\to0 $ at the end. 

After sending $x_5$ to infinity, as explained at the end of the previous section, we find that the Feynman integrals contributing to \re{Fintegrandfinal} and to the analogous expression for $F^{(3)}_{\rm planar}$ belong to a family of integrals with $15$ denominator factors  
\begin{equation}\label{definitionFamily}
\begin{aligned}
G_{a_1, \ldots , a_{15}} :=& \int \frac{ d^{D}x_{6} d^{D}x_{7} d^{D}x_{8}}{ (i \pi^{D/2})^3} 
\frac{(-1)^{\sum_{i=1}^{15} a_{i}}}{(x_{36}^2)^{a_1}  (x_{16}^2)^{a_2}  (x_{37}^2)^{a_3}  (x_{17}^2)^{a_4} (x_{38}^2)^{a_5}  (x_{18}^2)^{a_6}
} \\
& \hspace{-1cm}  \times \frac{1}{(x_{26}^2)^{a_7}  (x_{67}^2)^{a_8}  (x_{78}^2)^{a_9}  (x_{84}^2)^{a_{10}} (x_{64}^2)^{a_{11}}  (x_{72}^2)^{a_{12}}  (x_{74}^2)^{a_{13}}  (x_{82}^2)^{a_{14}}   (x_{68}^2)^{a_{15}}    } \,.
\end{aligned}
\end{equation}
We will see presently that this integral resembles momentum integrals computed previously for three-loop two-to-two scattering processes \cite{Henn:2013fah}. In order to do so, we interpret the $x$'s as dual variables, according to
  \begin{align}
 x_1 = 0\,, \quad x_2 = p_1 \,,\quad x_3 = p_1+p_2 \,, \quad x_4 = -p_4 \,, \quad  x_6=k_1 \,,\quad  x_7 = k_2 \,, \quad x_8=k_3 \,.  
 \end{align}
Here, we consider four $p_{i}$ corresponding to a four-particle scattering process. We take $p_{i}$ as ingoing, ordered according to 
$p_{1}, p_{2}, p_{3}, p_{4}$\footnote{Our notation for the integral family follows the one for the triple ladder integral in ref. \cite{Henn:2013fah}
as far as the labelling of propagators is concerned. However, note that here we use a different ordering of external momenta.}. 
Furthermore, the sum over momenta is conserved and they are light-like.
\beq
\sum\limits_{i=1}^{4} p_{i} =0,\hspace{1cm} p_i^2=0.
\eeq
We define scalar products of the momenta by
\bea
2 p_1 \cdot p_2= s =x_{13}^2\,,\qquad 
2 p_1 \cdot p_3=  t =x_{24}^2\,,\qquad 
2 p_2 \cdot p_3= -s-t\,.
\eea
Using the definition (\ref{definitionFamily}), one may represent all planar three-loop two-to-two scattering integrals. For example, the triple ladder integral  is given by $G_{1,1,1,1,1,1,1,1,1,1,0,0,0,0,0}$, see  Figure~\ref{figuregintegrals}(a).
However, the use of dual coordinates has further advantages. 
It allows us also to write down products of lower-loop integrals. 
For example, the product of a one-loop box integrals with a two-loop ladder integral is given by $G_{1,1,1,1,1,1,1,1,0,1,0,0,1,1,0}$.  

What integrals occur in $F$? At leading color and up to three loops, we have precisely planar three-loop integrals, and products of planar lower-loop integrals. 
Up to the unusual fact that products are considered, these integrals have a clear interpretation in terms of planar scattering processes.
The non-planar corrections to $F$ do not have such a simple interpretation, but they can still be expressed in our notation.
We found the graphical representation shown in Fig.~\ref{figuregintegrals}(b) useful.

Some readers might wonder wether the integrals under consideration are related to non-planar three-loop integrals considered in refs. \cite{Henn:2013nsa,Henn:2016jdu}. This is not the case. Moreover, there is one important difference, which is somewhat surprising. Although the integrals defined in eq. (\ref{definitionFamily}) cannot in general be drawn as planar two-to-two scattering diagrams, they do share important analytic properties with planar integrals. It turns out that their Feynman parametrization is positive definite if $s, t<0$, such that in this region the integrals are real-valued. Moreover, it can be seen from Fig.~\ref{figuregintegrals}(b) that the integrals only have cuts in the $s$- and $t$-channel, just like planar integrals. For these reasons we expect them to be free of $u$-channel cuts \cite{Gehrmann:2000zt}. This will be important when fixing boundary conditions for differential equations in section \ref{section:de}.

In practice, in order to analyze the integrals containing polynomials $Q_{3}$, $R_{1}$ and $R_{2}$ (see Eqs.~\re{Fintegrandfinal} and \re{P-fin}),   it turns out that we need integrals with up to thirteen propagators, i.e. thirteen non-negative indices $a_{i}$ in eq. (\ref{definitionFamily}). Note that this is three more, and hence considerably more complicated, compared to the integrals that were computed in \cite{Henn:2013fah}. 

We performed the reduction of all integrals needed using the program FIRE5 \cite{Smirnov:2014hma}.
As a result, we obtained $257$ master integrals, meaning that any integral that we need can be expressed in terms of the latter.\footnote{Computing the integrals, we observe some additional 
identities, at least to a certain order in the $\eps$ expansion. Therefore this basis may be overcomplete. This does not affect the validity
of our calculation.}
In the next section, we explain how we chose a particularly convenient integral basis.

\subsection{Integrand analysis via improved Baikov variables}
\label{section:improvedbaikov}

In the previous section we found that the integral family of eq. (\ref{definitionFamily}) is described by $257$ master integrals.
Before proceeding with their calculation, we will choose a convenient integral basis for the latter.
In order to do so, we rely on insights into the structure of Feynman integrals from \cite{ArkaniHamed:2010gh,Henn:2013pwa}.
The main idea is that the Feynman loop integrand contains the key information on the structure of the functions appearing
after carrying out the space-time integrations. In particular, the singularity structure of the integrand, and its residues, 
allow one to predict key features of the outcome of the integration, such as the transcendental weight properties of the answer.

In practice, following  \cite{ArkaniHamed:2010gh,Henn:2013pwa}, we consider loop integrands of the type appearing in eq. (\ref{definitionFamily}),
and search for integrands that are free of double poles, and whose leading singularities are normalized to be independent of the external kinematics.
In order to do so, we make use of the algorithm \cite{WasserMSc} (and an independent implementation of it) that is based on a particular parametrization of the integration variables, combined with partial fractioning. Although this algorithm has been applied in many similar situations, the present case is challenging due to the large number of propagator factors. For this reason we made certain improvements that we discuss below.
 
The first improvement is based on the observation that certain loop integrands can be written as {\it dlog} differential forms,
in such a way that the integrand is expressed as a single term \cite{Arkani-Hamed:2014via,Bern:2015ple}. The key example we need
is that of the one-loop on-shell box integrand,
\begin{align}\label{improvedBaikov}
\frac{x_{13}^2 x_{24}^2 d^{4}x_{6}}{x_{16}^2 x_{26}^2 x_{36}^2 x_{46}^2 } = 
\frac{s  t\,  d^{4}k}{ k^2 (k+p_{1})^2 (k+p_{1}+p_{2})^2 (k-p_{4})^2 } = 
\pm \frac{d\alpha_{1}}{\alpha_1 }  \frac{d\alpha_{2}}{\alpha_2} \frac{d\alpha_{3}}{\alpha_3}  \frac{d\alpha_{4}}{\alpha_4} 
\end{align}
Here the new variables $\alpha_{i}$ are closely related to the original propagators, 
\begin{align}
\alpha_{1} = \frac{k^2}{(k-k^{*}_{\pm})^2} \,,\quad \alpha_{2} = \frac{(k+p_{1})^2}{(k-k^{*}_{\pm})^2} 
\,,\quad \alpha_{3} = \frac{(k+p_{1}+p_{2})^2}{(k-k^{*}_{\pm})^2} 
\,,\quad \alpha_{4} = \frac{(k-p_{4})^2}{(k-k^{*}_{\pm})^2} 
\end{align}
where $k^{*}_{\pm}$ corresponds to the two solutions of the maximal cut condition.
We note that this is closely related in spirit, but different to the Baikov representation.
There, one chooses as new integration variables a set of propagator factors (and possibly irreducible scalar products), in
$D$ dimensions. Here, we consider the four-dimensional part of the loop integrand, and our new variables are normalized
by $(k-k^{*}_{\pm})^2$. As a result, the change of variables is rational in our case.

Let us now consider the integrand in eq. (\ref{definitionFamily}) for $a_{i} =1 $, and with an arbitrary numerator $P(x_{ij}^2)$.  The denominator is naturally written in the dual $x$ coordinates, and contains three factors of the type (\ref{improvedBaikov}).
Therefore we find it natural to perform this change of variable for each loop integration variable. 
We do so with $\alpha_i, \beta_i$ and $\gamma_i$ corresponding to $x_{6}, x_{7}$ and $x_{8}$, respectively.
The only new calculation w.r.t. (\ref{improvedBaikov}) we need to do is for factors of the type $x_{67}^2$.
The latter is a rational function of $\alpha_i, \beta_i, s ,t $.
As a result, we obtain for the integrand
\begin{align}\label{improvedBaikov2}
I =  \prod_{i=1}^{4}  \frac{d\alpha_i}{\alpha_i}  \frac{ d\beta_i}{\beta_i } \frac{d\gamma_i}{\gamma_i}  \frac{P}{x_{67}^2 x_{78}^2 x_{68}^2} \,.
\end{align}
where $P$ stands for a polynomial in the $x_{ij}^2$ variables.
We call this representation improved Baikov representation.

Our goal is to find as many numerators $P$ as possible for which the integrand is free of double poles and has constant maximal residues.
In principle we could start with a general ansatz for $P$, subject to certain power counting constraints, and compute residues. As already mentioned, this turns out to be prohibitive due to the size of the integrand. For this reason,we use a divide-and-conquer approach. First of all, most integrals we are interested in contain only a subset of the $15$ propagators, so that we can perform a dedicated analysis for each of them. Each propagator structure defines what we call an integral sector. 
Second, instead of running the algorithm directly for each of these integral sectors, we analyze the integrand on cuts. 
In this way, each run of the algorithm is faster, and we can  easily combine the constraints on $P$ arising from each cut. 
One of the main advantages of the improved Baikov representation (\ref{improvedBaikov2}) is that it makes taking cuts simple.
In practice, we find that only a subset of cuts is needed to obtain candidates for {\it dlog} integrands. 
Finally, once a candidate is found, we can test it using our algorithm (this is considerably less complicated than running the algorithm for the whole ansatz).

In this way, we arrive at a large set of {\it dlog} integrands. 
We complement them with information on uniform weight integrals from ref.~\cite{Henn:2013fah}. 
We then perform the integral reduction of the corresponding ($D$-dimensional) integrals, which allows us to relate the candidate integrals to a basis of master integrals. 
We then select a linearly independent set of candidate integrals, and choose this as our new integral basis.
In the next section we derive and solve the differential equations satisfied by these integrals.

\subsection{Analytic computation of the master integrals}
\label{section:de}

Here we discuss the computation of the three-loop integrals via differential equations.
We denote by $f$ the basis of integrals that was found with the methods discussed in the last section.
We find that the differential equations (for reviews, see \cite{Argeri:2007up,Henn:2014qga}) for this basis takes the form 
\begin{align}\label{debasis}
d\, f(s,t;\epsilon) = \epsilon \,d \left[ A_1 \log s + A_2 \log t + A_3 \log(s+t) + A_4 \log \frac{ 1+ i \sqrt{s/t}}{1-i \sqrt{s/t}} \right] f(s,t;\epsilon) \,,
\end{align}
where $d = ds \partial_s + dt \partial_t $ and $A_i$ are some matrices.
These equations are in the canonical form proposed in \cite{Henn:2013pwa}, as expected for uniform weight integrals.
Indeed, it is easy to read off what properties the perturbative solution in $\eps$ has, as we discuss presently.  
As one can always choose one overall scale (by dimensional analysis), we can set $t=-1$ without loss of generality.
In this way it is clear that the solution to (\ref{debasis}) can be written, to any order in $\eps$, in terms of iterated integrals of the alphabet 
\begin{align} \label{alphabet}
\{ x , 1+x , (1+i \sqrt{x})/(1-i \sqrt{x})\} \,,
\end{align}
where $x=s/t$.
This is slightly more complicated compared to the differential equations for the three-loop integrals derived in \cite{Henn:2013fah,Henn:2013nsa,Henn:2016jdu}, which are described by the alphabet  $\{x, 1+x \}$.
We note that it is possible to rewrite the alphabet (\ref{alphabet}) in terms of linear letters, with singularities corresponding to fourth roots of unity, $1,i,-1,-i$, simply by changing variables to $x= -z^2$. In this way, one can rewrite the solution in terms of Goncharov polylogarithms. 
As we will see below, while the additional letter $(1+i \sqrt{x})/(1-i \sqrt{x})$ is needed in individual Feynman integrals, remarkably, we find that it drops out of the results for $F$.  

Let us now solve eq. (\ref{debasis}) for fixed $t=-1$ and $x = s/t$. 
As discussed in section \ref{section:defs}, $f$ is real-valued for $x>0$, and 
we expect a branch cut along the negative real $x$-axis, so that one needs 
to add a small imaginary part to $x$ when analytically continuing to that region.
This is relevant when implementing the finiteness of the integrals at $u=0$, as
discussed above, which corresponds to $x=-1$.

Our strategy for solving the differential equation is a follows: we write down the general solution in terms of iterated integrals, with base point $x = 0^{+}$. Here the superscript `${+}$' indicates that we approach $0$ keeping $x>0$. This specification is necessary since
the Feynman integrals under consideration in general diverge as $x \to 0$, with the rate of divergence being controlled by the matrix $A_1 $. We parametrize the solution in terms of real-valued integration constants at $x=0^{+}$. 

Then, we analytically continue to negative $x$, and impose the finiteness condition at $x=-1$. This constraint, together with the real-valuedness of the integration constants, turns out to fix all integration constants, except for a trivial overall normalization.\footnote{Typically, one needs to expand to some order $\eps^{n+k}$, with $k>0$, in order to obtain all constraints at order $\eps^n$.} This useful feature that the integration constants can be obtained effortlessly from physical consistency conditions, has already been observed and stressed in a number of calculations, see e.g. \cite{Gehrmann:2000zt,Henn:2013fah,Chicherin:2018old}. Finally, we fix the overall normalization by explicitly evaluating one trivial integral in terms of Gamma functions.

In this way, we can obtain the analytic solution for all master integrals to any desired order in $\eps$.
We are interested in computing $F$, a finite quantity. On the other hand, typical on-shell three-loop integrals contain
nested soft and collinear divergences that lead to up to $1/\eps^6$ poles. As a consequence, we need to
expand the solution of the integrals up to finite order, which corresponds to up to transcendental weight $6$. 

With this we can compute all integrals contributing to $F(x)$ to three loops.
In particular, an important check is the finiteness of the answer.
Note that this property is highly non-trivial, since individual three-loop integrals have poles up to $1/\eps^6$.
We verify that this is the case for the planar contributions to $F$.

Proceeding to the non-planar contribution, we are particularly interested in the non-planar integral $Q_3$ and the Gram determinants $R_1$ and $R_2$, as discussed in the previous section. We found there that $Q_{3}$ by itself does not vanish in the $D$-dimensional OPE limit, but the linear combination of eq. (\ref{P-fin}) does.
Indeed, we find that $Q_{3}$ has a $1/\eps$ pole in dimensional regularization, while the combination appearing in eq. (\ref{P-fin}) is finite.
Moreover, we checked explicitly that the finite result is independent of the free parameter $d_1$. We take this as supporting evidence that our procedure is consistent and produces an unambiguous finite answer.

\section{Wilson loop with Lagrangian insertion at three loops}
\label{section:Fintegratedresults}
 
Here we present our novel three-loop results for $F(x)$ defined in eq. (\ref{defexpansionF}).
The tree-level, one-loop, and two-loop contributions to $F$ have been computed in  \cite{Alday:2011ga,Alday:2012hy,Alday:2013ip}.
The first non-planar corrections can appear in eq. (\ref{defexpansionF}) at three loops, i.e. at order $\alpha_s^4$. They are the main objectives of our work.

As a warm-up we begin by reproducing the lower-loop results.
This serves as a welcome validation of our procedure, and of the integrals computed in the previous section.
We give the formulas for completeness. 
We express them in terms of harmonic polylogarithms (HPL) \cite{Remiddi:1999ew}, since the latter
will be appropriate when giving the three-loop results below.
We have
\begin{align}
F^{(0)}(x) =& -\frac{1}{4} \,, \label{Ftree}\\
F^{(1)}(x) =& 
\frac{1}{8} H_{\text{0,0}}+\frac{\pi ^2}{16}
 \label{Foneloop}\,, \\
\label{Ftwoloop}
F^{(2)}(x) =&
-\frac{\pi ^2}{16} H_{\text{0,0}}-\frac{3}{16} H_{\text{0,0,0,0}}+\frac{\pi^2}{32}   H_{\text{0,-1}} 
 +\frac{1}{16} H_{\text{0,-1,0,0}}+\frac{\pi ^2}{32} \
H_{\text{-1,0}}
 \nonumber \\ &
+\frac{1}{16} H_{\text{-1,0,0,0}}-\frac{\pi ^2}{16} \
H_{\text{-1,-1}}-\frac{1}{8} H_{\text{-1,-1,0,0}}+\frac{\zeta _3}{16} \
H_0-\frac{\zeta _3}{8} H_{-1}-\frac{107 \pi ^4}{5760}
 \,.
\end{align}
Here we omitted the argument $x$ of the harmonic polylogarithms, for brevity.

Let us now give the new results at three loops.
The planar result is given by
\begin{align}\label{FthreeloopP}
F_{\rm planar}^{(3)}(x) =& 
\frac{323 \pi ^4}{11520} H_{\text{0,0}}+\frac{\zeta _3}{32} \
H_{\text{0,0,0}}+\frac{9 \pi ^2}{64} H_{\text{0,0,0,0}}+\frac{15}{32} \
H_{\text{0,0,0,0,0,0}}-\frac{\pi ^4}{64} H_{\text{-1,0}}
 \nonumber \\ & \hspace{-0cm}
 -\frac{\zeta _3}{16} \
H_{\text{-1,0,0}}
 -\frac{\pi ^2}{16} H_{\text{-1,0,0,0}}-\frac{3}{16} \
H_{\text{-1,0,0,0,0,0}}+\frac{43 \pi ^4}{1920} H_{\text{-1,-1}}-\frac{\zeta 
_3}{16} H_{\text{-1,-1,0}}
 \nonumber \\ &
 +\frac{\pi ^2}{16} \
H_{\text{-1,-1,0,0}}+\frac{3}{16} H_{\text{-1,-1,0,0,0,0}}+\frac{\zeta 
_3}{8} H_{\text{-1,-1,-1}}-\frac{\pi ^2}{32} \
H_{\text{-1,-1,-1,0}}
 \nonumber \\ &
 -\frac{1}{16} H_{\text{-1,-1,-1,0,0,0}}+\frac{\pi \
^2}{16} H_{\text{-1,-1,-1,-1}}+\frac{1}{8} \
H_{\text{-1,-1,-1,-1,0,0}}-\frac{\pi ^2}{32} \
H_{\text{-1,-1,-2}}
 \nonumber \\ &
-\frac{1}{16} H_{\text{-1,-1,-2,0,0}}-\frac{\zeta _3}{16} \
H_{\text{-1,-2}}+\frac{\pi ^2}{32} H_{\text{-1,-2,0}}+\frac{1}{8} \
H_{\text{-1,-2,0,0,0}}-\frac{\pi ^2}{32} H_{\text{-1,-2,-1}}
 \nonumber \\ &
 -\frac{1}{16} \
H_{\text{-1,-2,-1,0,0}}+\frac{\pi ^2}{16} H_{\text{-1,-3}}+\frac{1}{8} \
H_{\text{-1,-3,0,0}}-\frac{\zeta _3}{8} H_{\text{-2,0}}-\frac{11 \pi \
^2}{192} H_{\text{-2,0,0}}
 \nonumber \\ &
 -\frac{5}{32} H_{\text{-2,0,0,0,0}}-\frac{\zeta 
_3}{16} H_{\text{-2,-1}}+\frac{\pi ^2}{48} H_{\text{-2,-1,0}}+\frac{1}{16} \
H_{\text{-2,-1,0,0,0}}-\frac{\pi ^2}{32} H_{\text{-2,-1,-1}}
 \nonumber \\ &
 -\frac{1}{16} \
H_{\text{-2,-1,-1,0,0}}+\frac{\pi ^2}{16} H_{\text{-2,-2}}+\frac{1}{8} \
H_{\text{-2,-2,0,0}}-\frac{11 \pi ^2}{192} H_{\text{-3,0}}-\frac{5}{32} \
H_{\text{-3,0,0,0}}
 \nonumber \\ &
 +\frac{3 \pi ^2}{32} H_{\text{-3,-1}}+\frac{3}{16} \
H_{\text{-3,-1,0,0}}-\frac{3}{16} H_{\text{-4,0,0}}+\frac{5 \left(\zeta 
_3\right){}^2}{32}+\frac{\pi ^2 \zeta _3}{192} H_0-\frac{\zeta _5}{8} \
H_0
 \nonumber \\ &
 +\frac{7 \zeta _5}{16} H_{-1}+\frac{3 \zeta _3}{16} H_{-3}-\frac{197 \pi \
^4}{11520} H_{-2}-\frac{3 \pi ^2}{32} H_{-4}+\frac{617 \pi ^6}{96768}
\,.
\end{align}
The non-planar part $F_{\rm non-planar}^{(3)}$ has a novel feature.
In addition to the dependence on harmonic polylogarithms, it contains a rational dependence on $x$, as follows
\begin{align}\label{FthreeloopNP}
F_{\rm non-planar}^{(3)} = \frac{1}{1+x} F_{\rm non-planar,1}^{(3)} + F_{\rm non-planar,2}^{(3)}\,.
\end{align}
We find
\begin{align}
F_{\rm non-planar,1}^{(3)} =& 
\frac{5 \pi ^4}{384} H_{\text{0,0}}+\frac{3 \zeta _3}{4} H_{\text{0,0,0}}+\frac{\pi ^2}{32} H_{\text{0,0,0,0}}+\frac{9}{16}
   H_{\text{0,0,0,0,0,0}}-\frac{\pi ^4}{128} H_{\text{-1,0}}
    \nonumber \\ & \hspace{-1.5 cm}
    -\frac{3 \pi ^2}{32} H_{\text{-1,0,0,0}}-\frac{15}{16}
   H_{\text{-1,0,0,0,0,0}}-\frac{3 \pi ^2}{16} H_{\text{-2,0,0}}-\frac{3}{4} H_{\text{-2,0,0,0,0}}+\frac{\pi ^2}{32}
   H_{\text{-3,0}}  
   \nonumber \\ & \hspace{-1.5 cm}
   +\frac{3}{16} H_{\text{-3,0,0,0}}+\frac{3}{8} H_{\text{-4,0,0}}+\frac{5 \pi ^2 \zeta _3}{16} H_0+\frac{27 \zeta _5}{16}
   H_0-\frac{\pi ^4}{16} H_{-2}+\frac{3 \pi ^2}{16} H_{-4}+\frac{739 \pi ^6}{80640} \,,
\\[3mm]
F_{\rm non-planar,2}^{(3)} =& 
-\frac{5 \pi ^4}{384} H_{\text{0,0}}-\frac{3 \zeta _3}{4} H_{\text{0,0,0}}-\frac{\pi ^2}{32} H_{\text{0,0,0,0}}-\frac{9}{16}
   H_{\text{0,0,0,0,0,0}}-\frac{13 \pi ^4}{960} H_{\text{-1,0}}
      \nonumber \\ & \hspace{-1.5 cm}
      +\frac{3 \zeta _3}{8} H_{\text{-1,0,0}}-\frac{\pi ^2}{16}
   H_{\text{-1,0,0,0}}-\frac{3}{8} H_{\text{-1,0,0,0,0,0}}+\frac{9 \pi ^4}{128} H_{\text{-1,-1}}+\frac{9 \pi ^2}{32}
   H_{\text{-1,-1,0,0}}
      \nonumber \\ & \hspace{-1.5 cm}
      +\frac{27}{16} H_{\text{-1,-1,0,0,0,0}}+\frac{\pi ^2}{32} H_{\text{-1,-2,0}}+\frac{3}{16}
   H_{\text{-1,-2,0,0,0}}-\frac{3 \pi ^2}{32} H_{\text{-1,-3}}-\frac{3}{16} H_{\text{-1,-3,0,0}}
      \nonumber \\ & \hspace{-1.5 cm}
      +\frac{3 \zeta _3}{8}
   H_{\text{-2,0}}-\frac{3}{8} H_{\text{-2,0,0,0,0}}-\frac{3 \pi ^2}{32} H_{\text{-2,-2}}-\frac{3}{16} H_{\text{-2,-2,0,0}}-\frac{\pi ^2}{32}
   H_{\text{-3,0}}-\frac{3}{16} H_{\text{-3,0,0,0}}
      \nonumber \\ & \hspace{-1.5 cm}
      +\frac{33 \zeta_3^2}{16}-\frac{\pi ^2 \zeta _3}{8} H_0-\frac{9 \zeta _5}{8}
   H_0-\frac{\pi ^2 \zeta _3}{4} H_{-1}-\frac{3 \zeta _5}{16} H_{-1}+\frac{3 \pi ^4}{320} H_{-2}+\frac{11 \pi ^6}{11520}
 \,.
\end{align}
Let us discuss the structure of these results. 
First of all, we note that the expressions for $F^{(L)}$  
have uniform transcendental weight $2 L$. From the point of view
of the integrals we computed, this corresponds to the maximal 
possible weight. This feature is typical of calculations in $\mathcal{N}=4$ 
super Yang-Mills theory.
Second, we find that the first non-planar correction (unlike the planar
one) depends on more than one rational structure.
This corresponds to the fact that the non-planar integrand has two leading
singularities, namely $1$ and $1/(1+x)$, while the leading singularities of
the planar integrand are $x-$independent. This is reminiscent of 
the situation for non-planar three-loop two-to-two scattering amplitudes, 
where the same leading singularities occur \cite{Henn:2016jdu}. 

\section{Cusp anomalous dimension from integration over Lagrangian insertion}
\label{section:Lintegration}

The knowledge of $F(x)$ at $(L-1)$ loops allows one to obtain the cusp anomalous dimension at $L$ loops. 
The detailed relation was found in \cite{Alday:2013ip}, 
and we briefly reviewed it in section \ref{sec_correlators},
see eqs. (\ref{extractgammacusp}) and (\ref{rulefunctional-intro}).
The net result is that we need to perform a certain integral transformation of $F(x)$.
The action of the latter on a power of $x$ was given in eq. (\ref{rulefunctional-intro}), and we repeat 
it here for convenience,
\begin{align}\label{rulefunctional}
\mathcal{I}[x^p] = \frac{\sin (\pi p)}{\pi p} \,. 
\end{align}
In this section, we show how to apply the functional $\mathcal{I}$ to the terms appearing in $F(x)$.

In particular, we are interested in the scenario where the function $F(x)$ is a linear combination of 
harmonic poly-logarithms of the form of our results ${F}^{(n)}$. 
First, we observe that the r.h.s. of eq. (\ref{rulefunctional}) vanishes if $p$ is a  non-zero, positive integer. 
Consequently, only the first, $O(x^0)$ term of a convergent Taylor series expansion of a function would contribute to this integral. 
However, the functions we are interested in contain also logarithms $\log^a (x)$ {for positive integers $a$}.
We find that 
\bea
\label{eq:integrationrule}
\mathcal{I}\left[x^p \log^a(x)\right]&=&\lim\limits_{\xi\rightarrow 0}\frac{\partial^a}{\partial\xi^a}  \mathcal{I}\left[x^{p+\xi} \right]\nonumber\\
&=& \frac{ a!  }{p^{a}}  \sum\limits_{k=0}^a\frac{(-1)^{a+p-1}(\pi p)^{k-1}}{ k!} \sin\left(\frac{\pi k}{2}\right) \,.
\eea
The above expression is valid for positive integer $p$.
Next, we want to apply this result to the type of functions we are interested in.
Note, that a HPL that admits a convergent power series expansion for $x\in [0,1]$ can be written as
\begin{align}\label{eq:hpldef} 
H_{-m_n,\dots,-m_1} (x)  
\equiv H_{\underbrace{\scriptstyle 0,\dots,0}_{m_n-1},-1,\dots,\underbrace{\scriptstyle 0,\dots,0}_{m_1-1},-1}(x)
=(-1)^n\sum\limits_{j_n>\dots>j_1\ge 1}  \frac{\left(-x\right)^{j_n}}{j_n^{m_n}} \frac{1}{j_{n-1}^{m_{n-1}}}\dots  \frac{1}{j_1^{m_1}}\,.
\end{align}
Our integration rule of eq.~\eqref{eq:integrationrule} may now be easily applied to each term of such a series expansion. 
The integral over HPLs that are regular at $x=0$ consequently vanishes. 
If they are multiplied by powers of $\log(x)$ we  obtain a closed form expression by manipulation of the indices appearing in the definition of a HPL (eq.~\eqref{eq:hpldef}). With this we find the following two integration rules.
\bea
\mathcal{I}\left[\log^a(x)H_{-m_n,\dots,-m_1} (x) \right]&=&(-1)^{1-a}  a! \sum\limits_{k=0}^a \frac{\pi^{k-1}}{k!} \sin\left(\frac{\pi k}{2}\right)H_{-(a+1-k+m_n) ,\dots,-m_1} (-1)\,,\nonumber\\
\mathcal{I}\left[\frac{\log^a(x)}{1+x}H_{-m_n,\dots,-m_1} (x)\right] 
&=&\mathcal{I}[\log^a(x)H_{-m_n,\dots,-m_1} (x) ] 
\\&+& 
(-1)^{a}  a!  \sum\limits_{k=0}^a \frac{\pi^{k-1}}{k!} \sin\left(\frac{\pi k}{2}\right)H_{-(a+1-k),-m_n,\dots,-m_1} (-1)\nonumber \,.
\eea
To derive the result of the second integration rule above one can simply make use of the series representation of $1/(1+x)$ around $x=0$ and re-arrange summation indices.
Notice that the right-hand side of the above integration rules is given by linear combinations of HPLs with argument $-1$, which can be evaluated in terms of  multiple-zeta values using for example the package HPL~\cite{Maitre:2005uu}. 
Since the functions ${F}$ are of the form considered above we are now equipped to apply eq.~\eqref{extractgammacusp}. 

\section{The four-loop cusp anomalous dimension in $\mathcal{N}=4$ sYM and QCD}
\label{section:cusp}

Applying the method of the previous section we evaluate $\mathcal{I}[F^{(L)}]$ for $L=0,1,2,3$ and use
eq. (\ref{extractgammacusp}) in order to determine the cusp anomalous dimension in $\mathcal{N}=4$ super Yang-Mills.
We find
\begin{equation}\label{resultcuspSYM}
\begin{aligned}
 {\Gamma}^{\rm}_{\rm cusp}(\alpha_{s}, N) =& \,  \left( \frac{\alpha_s  N }{ \pi }  \right) - \frac{\pi^2}{12}  \left( \frac{\alpha_s N }{ \pi }\right)^2 + \frac{11 \pi^4}{720}  \left( \frac{\alpha_s  N }{ \pi }\right)^3 
  \\
 & \hspace{-0.5cm} -  \left( \frac{\alpha_s N  }{ \pi }\right)^4 \left[     \frac{73 \pi^6}{20160} + \frac{ \zeta_{3}^2}{8}  + \frac{1}{N^2} 
 \left(  \frac{31}{5040}\pi^6 + \frac{9 \zeta_3^2}{4} \right)  \right] + \cO(\alpha_s^5) \,.
\end{aligned}
\end{equation}
This is the full result to four loops for the cusp anomalous dimension in $\mathcal{N}=4$ super Yang-Mills, for a Wilson loop in the adjoint representation.\footnote{Note that another common definition in the literature is $\gamma_{K} = 2\,  {\Gamma}_{\rm cusp}$.}
The expression is valid in the supersymmetric $\overline{\rm DR}$ scheme. The quartic Casimir term, and equivalently the first subleading color term in
eq. (\ref{resultcuspSYM}), is independent of the scheme choice at four loops.
We note that the non-planar four-loop contribution is negative.

We use (\ref{resultcuspSYM}) to derive the quartic Casimir part for pure Yang-Mills theory, following \cite{Henn:2019rmi}.
Together with the known $n_{f}$-dependent terms, this completes the full four-loop cusp anomalous dimension in QCD.
We present the full result here, in the $\overline{\rm MS}$ scheme.
First, we recall the expressions up to three loops \cite{Korchemsky:1987wg,Moch:2004pa}:
\begin{align} \notag
\label{eq:K123L}
\Gamma^{\rm QCD}_{{\rm cusp},R}={}& C_R \bigg\{ \frac{\alpha_s}{\pi} 
\\ \nn
&
+  \left(\frac{\alpha_s}{\pi}\right)^2 \left[ 
	 C_A   \left(\frac{67}{36}-\frac{\pi ^2}{12}\right) 
	-\frac{5}{9}  n_f T_F 
 \right] 
 \\ \nn &
+   \left(\frac{\alpha_s}{\pi}\right)^3  \left[ 
  C_A^2\left(\frac{11 \zeta_3}{24}+\frac{245}{96}-\frac{67 \pi ^2}{216}+\frac{11 \pi ^4}{720}\right)  \right.
  + n_f T_F   C_F  \left(\zeta_3-\frac{55}{48}\right)
\\ \nn 
&{} \phantom{\left(\frac{\alpha_s}{\pi}\right)^3 CA^3} + \left.  n_f T_F  C_A \left(-\frac{7 \zeta_3}{6}-\frac{209}{216}+\frac{5 \pi ^2}{54}\right)
-\frac{1}{27} (n_f T_F)^2 
\right] \bigg\} 
\nn
\\
+ &{} \left(\frac{\alpha_s}{\pi}\right)^4 \Gamma^{\rm QCD, (4)}_{{\rm cusp},R} 
+ \mathcal{O}(\alpha_s^5)
\,. 
\end{align}
At four loops, depending on the color factor, various terms were known either  analytically or numerically  \cite{Beneke:1995pq,Henn:2016men,Davies:2016jie,Lee:2016ixa,Moch:2017uml,Moch:2018wjh,Lee:2019zop,Henn:2019rmi,Bruser:2019auj,vonManteuffel:2019wbj,Gracey:1994nn,Grozin:2018vdn}.\footnote{Strictly speaking, the $n_{f} T_{F} C_{R} C_{F} C_{A}$ term is based on a conjecture \cite{Bruser:2019auj}. The latter agrees perfectly with the numerical value from \cite{Moch:2017uml,Moch:2018wjh}.} 
We present for the first time the fully analytic result, which is
\begin{align}\nn\label{eq:K4L}
\Gamma^{\rm QCD,(4)}_{{\rm cusp},R}={}& C_R\bigg[
  C_A^3 \left( \frac{1309 \zeta_{3}}{432}-\frac{11 \pi ^2 \zeta_{3}}{144}-\frac{\zeta_{3}^2}{16}-\frac{451 \zeta_{5}}{288}+\frac{42139}{10368}-\frac{5525 \pi^2}{7776}+\frac{451 \pi ^4}{5760}-\frac{313 \pi ^6}{90720} \right) 
  \nn\\&
+ n_f T_F  C_A^2 \left( -\frac{361 \zeta_{3}}{54}+\frac{7 \pi ^2 \zeta_{3}}{36}+\frac{131 \zeta_{5}}{72} -\frac{24137}{10368}+\frac{635 \pi ^2}{1944}-\frac{11 \pi ^4}{2160}  \right)  
  \nn\\&
 +  n_f T_F   C_F C_A \left(  \frac{29 \zeta_{3}}{9}-\frac{\pi ^2 \zeta_{3}}{6}+\frac{5 \zeta_{5}}{4}-\frac{17033}{5184}+\frac{55 \pi ^2}{288}-\frac{11 \pi^4}{720} \right)  
  \nn\\&
 + n_f T_F  C_F^2 \left( \frac{37 \zeta _3}{24}-\frac{5 \zeta _5}{2}+\frac{143}{288} \right) \nn
 + (n_f T_F)^2   C_A \left( \frac{35 \zeta _3}{27}-\frac{7 \pi ^4}{1080}-\frac{19 \pi ^2}{972}+\frac{923}{5184} \right)    
  \nn\\&
+ (n_f T_F)^2   C_F \left(-\frac{10 \zeta _3}{9}+\frac{\pi ^4}{180}+\frac{299}{648}\right) 
+ (n_f T_F)^3  \left(-\frac{1}{81}+\frac{2 \zeta _3}{27}\right)
 \bigg]  \nn
 \\
+  &{} 
  \frac{d^{abcd}_R d^{abcd}_A}{N_R} \left(  \frac{\zeta_{3}}{6}-\frac{3 \zeta_{3}^2}{2}+\frac{55 \zeta_{5}}{12}-\frac{\pi^2}{12}-\frac{31 \pi ^6}{7560}   \right)
  +n_f \frac{d^{abcd}_R d^{abcd}_F}{N_R} \left( \frac{\pi^2}{6}-\frac{\zeta_3}{3}-\frac{5\zeta_5}{3} \right) 
\,.  
\end{align}
For convenience of the reader we also print the above formula to six significant digits. 
\begin{align}
\Gamma^{\rm QCD}_{{\rm cusp},R}{}=& \, C_R \bigg[ \left(\frac{\alpha_s}{\pi} \right)  \nonumber\\
&+ \left( \frac{\alpha_s}{\pi} \right)^2  \left(1.03864 C_A -0.555556  n_f T_F\right)\nonumber\\
&+\left( \frac{\alpha_s}{\pi} \right)^3  \left(
	 1.52982 C_A^2
	 -1.45614 C_A n_f T_F
	 +0.0562236 C_F n_f T_F
	 -0.0370370 n_f^2 T_F^2
 \right)\nonumber\\
&+\left( \frac{\alpha_s}{\pi} \right)^4  \left(
	2.38379 C_A^3
	-3.44271 C_A^2 n_f T_F
	+0.303089 C_A C_F n_f T_F
	\right.\nonumber\\
& \quad \left.
	-0.242621 C_F^2 n_f T_F
	+0.911990 C_A n_f^2 T_F^2
	-0.333037 C_F n_f^2 T_F^2
	+0.0766956 n_f^3 T_F^3 
 \right) \bigg] \nonumber\\
&+\left( \frac{\alpha_s}{\pi} \right)^4 \left(
 -\frac{d^{abcd}_R d^{abcd}_A}{N_R}  1.97915  
 -n_f \frac{d^{abcd}_R d^{abcd}_F}{N_R} 0.483964  
  \right) \,.
\end{align}
This result is given for an arbitrary representation $R$ of the Wilson loop.
In the QCD context there two choices for the representation, namely $R=A$ (gluons) and $R=F$ (quarks).
For the normalization of the $SU(N)$ generators, we follow the conventions of \cite{vanRitbergen:1997va,vanRitbergen:1998pn},
which are
\begin{align}
  \frac{d_{A}^{abcd}d_{A}^{abcd}}{N_{A}} &= \frac{N^2 (N^2+36)}{24} \,,\quad
 \frac{d_{F}^{abcd}d_{A}^{abcd}}{N_{A}} =\frac{N (N^2+6)}{48} \,,\quad    \frac{d_{F}^{abcd}d_{F}^{abcd}}{N_{A}} = \frac{N^4-6 N^2+18}{96 N^2} \,,\nn\\
 T_{F}& = \frac{1}{2} \,,\quad C_{A} = N \,,\quad C_{F} =\frac{N^2-1}{2 N} \,,\quad N_{A} = N^2-1 \,,\quad N_{F} = N .
\end{align}
We remark that if one retains only the leading transcendental pieces of the QCD result (\ref{eq:K4L}), i.e. the transcendental weight $2 (L-1)$ pieces at $L$ loops, and switches to the adjoint representation $R\to A$, then one recovers (\ref{resultcuspSYM}), as expected \cite{Kotikov:2004er,Dixon:2017nat}.

\section{Conclusion and outlook}
\label{section:conclusion}

In this paper we computed analytically the non-planar part of the cusp anomalous dimension in 
$\mathcal{N}=4$ super Yang-Mills at four loop. Our result agrees with the previous numerical
results of refs.~\cite{Boels:2017skl,Henn:2019rmi}. 
$\mathcal{N}=4$ super Yang-Mills is expected to be integrable, and there is considerable interest
in finding ways to solve the theory. The planar part of the cusp anomalous dimension is 
successfully described by integrability \cite{Beisert:2006ez}, with important input from
perturbative calculations \cite{Bern:2006ew}. We expect that our result, which constitutes
the leading non-planar contribution to this quantity, will be an important reference value for
future integrability studies.

As was mentioned in the Introduction, the cusp anomalous dimension controls behavior of the DGLAP splitting functions close to the end-point, or equivalently the
large spin asymptotics  of twist-two anomalous dimensions, $\gamma(S) = 2\Gamma_{\rm cusp} \log S + O(S^0)$  \cite{Korchemsky:1988si}. Using the obtained result, we can predict nonplanar correction to these anomalous dimensions. At present, nonplanar corrections to $\gamma(S)$ are known for lowest value of spins $S=0,2,4,6,8$ both in $\mathcal N=4$ sYM and in QCD \cite{Velizhanin:2009gv,Velizhanin:2010ey,Velizhanin:2014zla,Fleury:2019ydf}. These expressions exhibit an interesting structure and several conjectures have been formulated about the possible form of nonplanar corrections to $\gamma(S)$ for arbitrary spin $S$. 
It would be very interesting to find such a formula. Our result provides the large spin asymptotics of $\gamma(S)$  
and it can be used to constrain an ansatz for this function.
 
Having the result in $\mathcal{N}=4$ super Yang-Mills, we derived the value for the purely gluonic
quartic Casimir contribution to the cusp anomalous dimension in QCD. The analytic value we
computed agrees with the numerical result of ref.~\cite{Moch:2018wjh}. Our result provides the last 
missing ingredient for the full four-loop result in QCD. By assembling the known terms from the literature, 
together with our new result for the gluonic quartic Casimir term, we presented for the first time
the complete result for the four-loop cusp anomalous dimension in QCD, for an arbitrary color representation
of the Wilson lines.

For convenience of the reader we provide ancillary files with our paper that contain the results for $F(x)$ given in eqs. (\ref{Ftree}) - (\ref{FthreeloopNP}),
as well as the result for the cusp anomalous dimension in $\mathcal{N}=4$ sYM, eq. (\ref{resultcuspSYM}), and QCD, eqs. (\ref{eq:K123L}) and (\ref{eq:K4L}).

\section*{Acknowledgments}
We thank P.~Wasser for discussions on integrands with logarithmic differential forms and for help in using the program \cite{WasserMSc},
and M.~Stahlhofen for discussions.
J.M.H. thanks the staff at ICTP-SAIFR for hospitality during the beginning of this project. J.M.H. and B.M.
thank KITP Santa Barbara for hospitality during the program {\it Scattering Amplitudes and Beyond}.
G.K. thanks the Max-Planck-Institute for Physics for hospitality during the final stages of this project.
This  research  received  funding  from  the  European  Research  Council  (ERC)  under  the 
European Unions Horizon 2020 research and innovation programme (grant agreement No725110),
{\it Novel structures in scattering amplitudes}.  B.M. is supported by a Pappalardo fellowship and G.K.\ by the French National Agency for Research grant ANR-17-CE31-0001-01.

\bibliographystyle{JHEP}
\bibliography{bibfile_w4p1.bib}

\end{document}